\documentclass[]{spie}  

\usepackage{amsfonts}
\usepackage{amsmath}

\usepackage{amsmath,amsfonts,amssymb}
\usepackage{graphicx}
\usepackage[colorlinks=true, allcolors=blue]{hyperref}

\title{Synthesizing Audio from Tongue Motion During Speech Using Tagged MRI Via Transformer}

\author[a]{Xiaofeng Liu}
\author[a]{Fangxu Xing}
\author[b]{Jerry L. Prince}
\author[c]{Maureen Stone}
\author[a]{Georges El Fakhri}
\author[a]{Jonghye Woo}

\affil[a]{{Gordon Center for Medical Imaging, Massachusetts
General Hospital and Harvard Medical School, Boston, MA 02114 USA}}
\affil[b]{Department of Electrical and Computer Engineering, Johns Hopkins University, Baltimore, MD 21218 USA}
\affil[c]{{Department of Neural and Pain Sciences, University of Maryland School of Dentistry, Baltimore, MD 21201 USA}}

\begin{document} 
\maketitle

\vspace{+5pt}
\begin{abstract}

Investigating the relationship between internal tissue point motion of the tongue and oropharyngeal muscle deformation measured from tagged MRI and intelligible speech can aid in advancing speech motor control theories and developing novel treatment methods for speech related-disorders. However, elucidating the relationship between these two sources of information is challenging, due in part to the disparity in data structure between spatiotemporal motion fields (i.e., 4D motion fields) and one-dimensional audio waveforms. In this work, we present an efficient encoder-decoder translation network for exploring the predictive information inherent in 4D motion fields via 2D spectrograms as a surrogate of the audio data. Specifically, our encoder is based on 3D convolutional spatial modeling and transformer-based temporal modeling. The extracted features are processed by an asymmetric 2D convolution decoder to generate spectrograms that correspond to 4D motion fields. Furthermore, we incorporate a generative adversarial training approach into our framework to further improve synthesis quality on our generated spectrograms. We experiment on 63 paired motion field sequences and speech waveforms, demonstrating that our framework enables the generation of clear audio waveforms from a sequence of motion fields. Thus, our framework has the potential to improve our understanding of the relationship between these two modalities and inform the development of treatments for speech disorders.

\vspace{+5pt}
\end{abstract}


\vspace{+5pt}
\section{Introduction}\vspace{+5pt}

To advance our understanding of speech motor control in both healthy and diseased populations, such as tongue cancer patients, it is important to identify the relationships between dynamic magnetic resonance imaging (MRI) data and speech audio waveforms. This can help us associate tongue and oropharyngeal muscle deformation with its corresponding acoustic information. Internal tissue point tracking data from three-dimensional (3D) tagged MRI~\cite{woo2021deep} sequences contain far more information about the tongue and oropharyngeal motion than does the more conventional two-dimensional (2D) mid-sagittal image sequences obtained from cine-MRI~\cite{liu2022cmri2spec} and tagged MRI~\cite{liu2022tagged}. Yet, associating these four-dimensional (4D) deformation fields with speech audio waveforms poses the following challenges:~1) efficient feature extraction from complex and high-dimensional tongue and oropharyngeal deformation and 2) heterogeneous data representations between 4D motion fields and high-frequency one-dimensional (1D) audio waveforms.

To tackle these challenges, we present a novel framework for synthesizing a 2D Mel-spectrogram from 4D motion fields using an efficient heterogeneous translation framework. We utilize 2D spectrograms as a proxy representation, a representation commonly used in audio-visual translation tasks, which is obtained by converting the 1D audio waveform in this work, as in~\cite{liu2022cmri2spec}. 
Previous research on the translation of 2D MRI sequences to audio~\cite{liu2022cmri2spec,liu2022tagged} has demonstrated that the 2D spectrogram is an effective representation for this task, as it captures the distribution of acoustic energy across frequencies over time~\cite{akbari2018lip2audspec,ephrat2017vid2speech,he2020image2audio,he2020classification}. To exploit the rich spatiotemporal information in 4D motion fields, we propose a novel efficient encoder network, involving a combination of a 3D convolutional neural network (CNN) and a Longformer-based transformer module to synthesize spectrograms from the motion fields. Specifically, we apply a 3D CNN for the spatial modeling of motion fields at each time point, followed by applying a Longformer \cite{beltagy2020longformer}-based transformer module for temporal modeling. Compared with conventional temporal modeling methods used in 2D MRI sequence processing, e.g., recursive neural networks (RNN) and 3D CNN~\cite{akbari2018lip2audspec,liu2022cmri2spec,liu2022tagged}, our transformer module takes more than 1.5$\times$ fewer parameters and can be trained on fewer training samples than conventional approaches. Then, a 2D convolutional generator is applied to yield spectrograms, which can then be converted back into the corresponding audio waveforms \cite{griffin1984signal}. We also incorporate generative adversarial training to further improve the quality of the synthesized spectrograms. Our framework represents the first attempt at learning the mapping between 4D motion fields and audio waveforms and offers the potential to better understand the relationship between motion and intelligible speech. 

\begin{figure}[t]
\begin{center}
\includegraphics[width=1\linewidth]{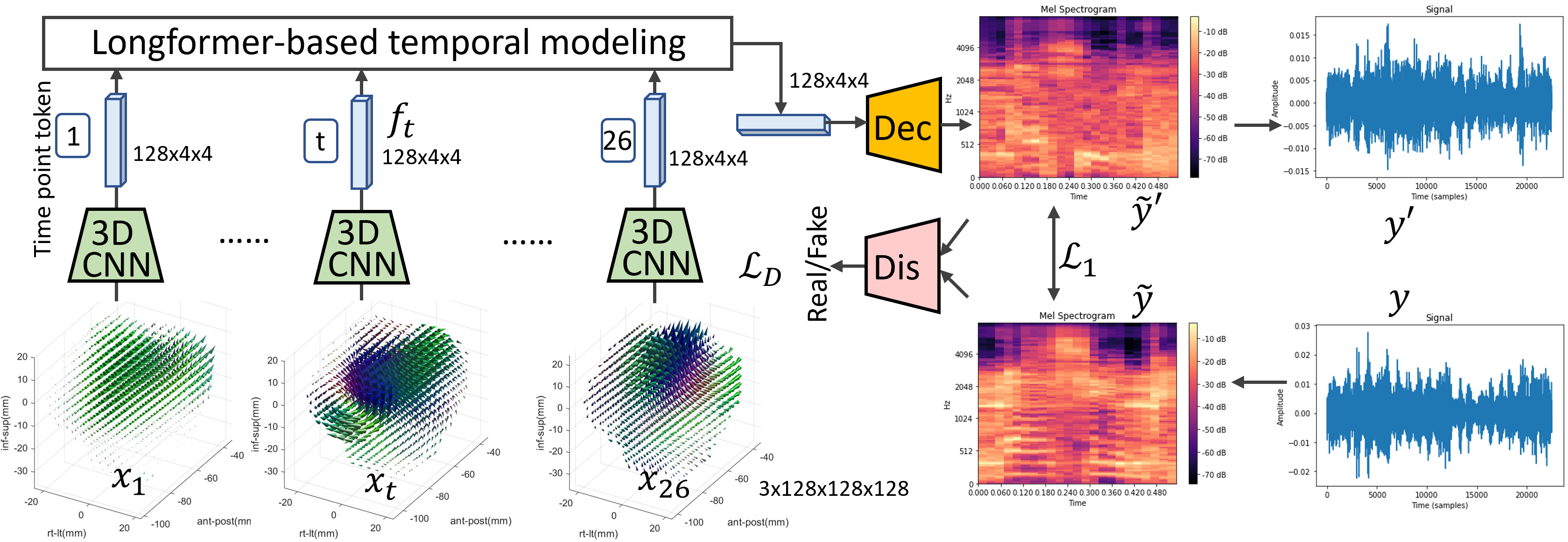}
\end{center} 
\caption{Illustration of our framework for synthesizing audio waveforms from a sequence of motion fields, which consists of a two-stage encoder (with 3D CNN and Longformer), a 2D convolutional decoder (Dec), and a discriminator (Dis).} 
\label{ccc}\end{figure}

\vspace{+5pt}
\section{METHODS}\vspace{+5pt}

We are given a set of motion fields $x$ with the size of ${3\times N\times H\times W\times T}$ along with its corresponding 1D waveform $y$, where $N, H, W$, and $T$ denote the mid-sagittal slice number, height, width, and time frame number, respectively. It is worth noting that each voxel of the motion fields has three channels to represent 3D directions, whereas cine or tagged-MRI sequences have a simpler data structure of ${H\times W\times T}$~\cite{liu2022cmri2spec,liu2022tagged}. Each audio waveform $y$ is pre-processed into a 2D Mel-spectrogram $\tilde{y}\in\mathbb{R}^{64\times64}$ using Librosa\footnote{\href{https://librosa.org/doc/main/generated/librosa.feature.melspectrogram.html.}{Librosa: generating mel-spectrogram from audio waveforms.}}. The Mel-spectrogram uses the mel-scale, a non-linear transformation of the Hz-scale, to emphasize human voice frequencies from 40 to 1000 Hz and suppress high-frequency instrument noise. The goal of this work is to learn an end-to-end heterogeneous translator $\mathcal{T}:x\rightarrow\tilde{y}'$ that approximates $\tilde{y}$.

To make use of the rich information in $x$, we adopt a modular design for spatial and temporal information modeling, similar to~\cite{neimark2021video}. First, a 3D CNN module is applied to the three-channel 3D motion field at each time point $t\in\{1,\cdots,T\}$ to extract a compact representation feature $f_t\in\mathbb{R}^{128\times4\times4}$. The detailed 3D CNN for each motion field is shown in Table 1. In previous work on video-to-audio synthesis, RNNs and 3D CNNs have been widely used for temporal modeling. However, both RNNs and 3D CNNs, when applied to temporal modeling, have their own challenges, including difficulty in training RNNs on limited datasets~\cite{liu2022tagged} and difficulty in modeling long-term correlations with 3D CNN, respectively~\cite{wang2021automated}. 

Inspired by recent developments in vision transformers~\cite{neimark2021video}, we propose using a transformer module that applies attention mechanisms to explore global dependencies within a sequence. While vanilla transformers can only process pairwise correlations of limited tokens or time points and are not scalable to long sequences, the recent development of Longformer~\cite{beltagy2020longformer} with sliding window attention has linear complexity with respect to the length of the sequence. As in {Ref.~\cite{neimark2021video}}, the Longformer module takes both the feature at each time point and the time point index $t$ to fuse the information and generate a sequence-level representation. It is worth noting that the attention scheme used in the transformer is permutation invariant, so the time point index is essential for embedding sequential information. We use three Longformer layers as our temporal transformer module. The processing flow is shown in Fig. 1. The parameters of the Longformer module are 1.5$\times$ fewer than those of the 3D CNN-based temporal modeling\cite{neimark2021video}, making it a lighter network that may outperform the 3D CNN with relatively limited data sets. In contrast to the conventional 3D CNN-based temporal modeling~\cite{neimark2021video}, which can only focus on neighboring frames for short-term temporal modeling, the transformer module is able to model long-term temporal relationships, potentially contributing to more representative audio features. After generating a global representation of the 4D motion fields, we use a 2D decoder to render the 2D spectrogram $\tilde{y}$, which is compared to the ground truth $\tilde{y}$ using the L1 loss, $\mathcal{L}_1$.

\begin{table}[t]
\centering\caption{Structure of the proposed networks for synthesizing audio from tongue motion during speech using Tagged MRI via transformer}  \vspace{+8pt}
\resizebox{1\linewidth}{!}{
  \centering
  \begin{tabular}{ccccc}
    \hline 
    \multicolumn{2}{c}{Encoder (3D CNN)+Longformer}  &  & \multicolumn{2}{c}{Decoder}  \\
     \cline{1-2} \cline{4-5}
    Layers     & Size  & & Layers     & Size  \\
    \cline{1-2} \cline{4-5}
    Input  & (3, 128, 128, 128)$\times 26$  &&  Reshape & (128, 4, 4)   \\
    \cline{1-2} \cline{4-5}
    Conv3D (32) \& ReLU  &(32, 128, 128)$\times 26$&& Conv2DTrans(96) \& ReLU &(96, 8, 8) \\
     
    MaxPooling &(32, 64, 64)$\times 26$  \\
	\cline{1-2} \cline{4-5} 
	
    Conv3D (32) \& ReLU &(32, 64, 64)$\times 26$ && Conv2DTrans(24) \& ReLU &(24, 16, 16) \\
   
    MaxPooling & (32, 32, 32)$\times 26$   \\
	\cline{1-2} \cline{4-5}   
	
    Conv3D(64) \& ReLU &(64, 32, 32)$\times 26$&& Conv2DTrans(4) \& ReLU  &(4, 32, 32)\\
  
    MaxPooling &(64, 16, 16)$\times 26$   \\
	\cline{1-2} \cline{4-5}
	
    Conv3D (64) \& ReLU &(64, 16, 16)$\times 26$&& Conv2DTrans(1) \&  sigmoid  &(1, 64, 64) \\
    
    MaxPooling &  (64, 8, 8)$\times 26$ \\
	\cline{1-2} \cline{4-5} 
	
    Conv3D (128) \& ReLU &(128, 8, 8)$\times 26$&&   \\
    MaxPooling & ({128}, 4, 4)$\times 26$ & & Librosa to audio waveform\\
	\cline{1-2} \cline{4-5} 
    Longformer (3 layers, sliding window size of 3) &(128, 4, 4) &&   \\
  
    \hline 
  \end{tabular}
}
\label{table:network-table} 
\end{table}

We also include a generative adversarial network (GAN) module to further improve the quality of our generated Mel-spectrograms. The discriminator $\mathcal{D}$ takes as input both the real spectrogram ${\tilde{y}}$ and the generated spectrogram $\tilde{{y}}'$, and is tasked with identifying which is generated and which is real. The binary cross-entropy loss of the discriminator can be expressed as \vspace{+3pt}
\begin{equation}
    \mathcal{L}_{\mathcal{D}} = \mathbb{E}_{{y}'}\{\text{log}(\mathcal{D}({y}'))\} +  \mathbb{E}_{\tilde{{y}}'}\{\text{log}(1-\mathcal{D}(\tilde{{y}}'))\}.
\end{equation}\vspace{+3pt}
In contrast, the translator tries to fool the discriminator by generating realistic spectrograms~\cite{liu2021dual}. It is worth noting that the translator $(\mathcal{T})$ does not involve real spectrograms in $\text{log}(\mathcal{D}({y}'))$~\cite{salimans2016improved}. As a result, the translator can be trained by optimizing \vspace{+3pt}
\begin{equation}
\mathcal{L}_{\mathcal{T}}= \mathbb{E}_{\tilde{{y}}'}\{-\text{log}(1-\mathcal{D}(\tilde{{y}}'))\}+\beta\mathcal{L}_1.
\end{equation}\vspace{+3pt}
After the training stage, $\tilde{y}'$ can be converted back into the audio waveform $y'$\footnote{\href{https://librosa.org/doc/main/generated/librosa.feature.inverse.mel_to_audio.html}{Librosa: generating audio waveforms from Mel-spectrogram.}}.

\vspace{+5pt}
\section{RESULTS}\vspace{+5pt}

To evaluate our framework, we collected a dataset of paired MRI sequences and audios with a Siemens 3.0T TIM Trio system. Our collected data consist of a total of 43 subjects who performed ``a geese," and a total of 20 subjects who performed ``a souk," following a periodic metronome-like sound. We then computed a sequence of voxel-level motion fields during the speech tasks from tagged MRI~\cite{xing2017phase,woo2021deep}. The data for this work were collected using a Siemens 3.0T TIM Trio system equipped with a 12-channel head coil and a 4-channel neck coil, using a segmented gradient echo sequence~\cite{lee2013semi,xing20133d}. The 4D motion fields $x$ has the size of ${3\times128\times128\times128\times26}$. In contrast, the length of the paired 1D audio recordings in our dataset ranges from 21,832 to 24,175 samples. We adopted a sliding window to crop the audio waveform to a length of 21,000, generating 100$\times$ audio waveforms for data augmentation. Then, we used the publicly available Librosa library to convert the audio waveforms into Mel-spectrograms with the size of $64\times64$. For testing, we used a leave-one-out evaluation in a subject-independent manner.

Our framework was implemented using PyTorch and was trained on an NVIDIA V100 GPU. For Longformer, we used an attention window of three frames, which was applied to each layer. We set the momentum to 0.5 and the learning rate of the encoder-decoder and discriminator to $10^{-3}$ and $10^{-4}$, respectively. The loss term $\mathcal{L}_{\mathcal{T}}$ was balanced using $\beta=1$. In testing, the inference time for one subject was less than 0.5 seconds. We applied the proposed 3D CNN module to the motion fields at each time frame, followed by applying either the 3D CNN or the Longformer-based transformer module for temporal modeling, which are denoted as a two-stage 3D CNN or a transformer, respectively.

An example of the predicted spectrogram and audio waveform is provided in Fig. 1, demonstrating that the audio can be generated from a sequence of motion fields. To quantify the quality of the generated spectrogram in the frequency domain and audio waveform in the time domain, we used the 2D Pearson's correlation coefficient (Corr2D) and Perceptual Evaluation of Speech Quality (PESQ) as in Refs.~\cite{akbari2018lip2audspec,liu2022tagged}, respectively. Higher values of Corr2D and PESQ indicate better synthesis performance. The numerical comparison results, including standard deviations from three random trials, are shown in Table 2. Our proposed transformer framework, comprising a 3D CNN and Longformer, achieved superior performance on both Corr2D and PESQ metrics.

The training time for 200 epochs and inference time in testing were 1.7$\times$ and 1.3$\times$ faster, respectively, compared with the two-stage 3D CNN. An ablation study was conducted to test the effectiveness of the GAN loss, which was found to improve performance. Sensitivity analysis of the parameter $\beta$, which balances the GAN loss and L1 loss, showed that system performance was relatively stable for $\beta\in[0.5,7]$.

\vspace{+5pt}
\section{CONCLUSION}\vspace{+5pt}


In this work, we presented a novel synthesis framework that translates a sequence of motion fields into a corresponding spectrogram. Our modular, two-stage framework combines 3D CNN-based spatial information modeling with transformer-based temporal modeling to effectively utilize the small training set and complex structure of 4D motion fields. We also successfully applied adversarial training to further enhance performance. Our experiments demonstrated that our framework was able to generate spectrograms and intelligible audio from 4D motion fields, outperforming the 3D CNN when it comes to temporal modeling. This framework could potentially be adapted for other tasks, involving the translation of heterogeneous temporal sequences.



     
\begin{table}[t]
\centering 
\caption{Numerical comparisons in testing with leave-one-out evaluation. The best results are \textbf{bold}.} \vspace{+8pt}
\resizebox{1\linewidth}{!}{
\begin{tabular}{l|c|c|c}
\hline
Methods& ~~with GAN~~ & ~~~Corr2D for spectrogram $\uparrow$~~~ & ~~~PESQ for waveform $\uparrow$~~~\\\hline\hline
3D CNN + Transformer& $\surd$ & \textbf{0.820}$\pm$0.017  &  \textbf{1.646}$\pm$0.019  \\ 
3D CNN + Transformer & $\times$	&  {0.818}$\pm$0.015  &   1.632$\pm$0.022 \\\hline
Two-stage 3D CNN  &  $\surd$ 	   &{0.812}$\pm$0.018  &  1.630$\pm$0.020\\\hline
Two-stage 3D CNN  & $\times$	   &{0.807}$\pm$0.021  &  1.625$\pm$0.017\\\hline
\end{tabular}} 
\label{tabel:1}  
\end{table}

\vspace{+5pt}
\acknowledgments 

This work is supported by NIH R01DC014717, R01DC018511, and R01CA133015.


\bibliography{main} 
\bibliographystyle{spiebib} 

\end{document}